\documentclass[12pt]{s-article}
\usepackage{mathtools,amsfonts,amssymb,stackrel,bm,mathrsfs,mathabx,nicefrac}
\usepackage{epsfig,tikz,graphicx,amssymb,epstopdf,empheq}
\usepackage{array,adjustbox,multirow,xcolor,bigints}
\usepackage[parfill]{parskip}  
\usepackage[utf8]{inputenc}
\usepackage{geometry,caption}
\graphicspath{{./figures/}}

\newcommand{\ie}{{\it i.e.}}

\newcommand{\missing}[1]{{\it [$\rightarrow$~#1]}}

\newcommand{\qqwhere}{\qquad \text{where}\qquad}

\newcommand{\qqand}{\qquad \text{and}\qquad}

\newcommand{\vcr}[1]{\bm{\mathrm {#1}}}  
\newcommand{\vcrg}[1]{\boldsymbol {#1}}  

\newcommand{\vcf}[1]{\pmb{#1}}                 

\newcommand{\ee}[1]{\text{e}^{#1}}
\newcommand{\expb}[1]{\exp\left\{ #1 \right\}}

\newcommand{\sst}[2]{{#1}_{\text{#2}}}

\newcommand{\breathe}[2]{\rule[-#1ex]{0cm}{#2ex}}

\newcommand{\dx}{\text{d}x}
\newcommand{\dy}{\text{d}y}

\newcommand{\pder}[2]{\frac{\partial #1}{\partial #2}}

\newcommand{\pderh}[3]{\frac{\partial^{#3} #1}{\partial {#2}^{#3}}}

\newcommand{\psii}{\sst{\psi}{i}}

\newcommand{\psis}{\sst{\psi}{s}}

\newcommand{\lonenorm}[1]{\left|\!\left|#1\right|\!\right|_1}

\newcommand{\argmin}[1]{\stackrel[#1]{}{\text{argmin}}}

\newcommand\centergraphics[2]{\begin{center}
\includegraphics[width=#1\linewidth]{#2}
\end{center}}

\newcommand\balphlist[1]{\newcounter{#1}
\begin{list}
  {{\bf \alph{#1})}}   
  {\usecounter{#1}\setlength{\rightmargin}{0cm}
   \itemsep 1ex\parsep 1ex}}
\newcommand\palphlist[1]{\newcounter{#1}
	\begin{list}
		{(\alph{#1})}   
		{\usecounter{#1}\setlength{\rightmargin}{0cm}
			\itemsep 1ex\parsep 1ex}}
\newcommand\bpalphlist[1]{\newcounter{#1}
	\begin{list}
		{{\bf (\alph{#1})}}   
		{\usecounter{#1}\setlength{\rightmargin}{0cm}
			\itemsep 1ex\parsep 1ex}}
\newcommand\alphlist[1]{\newcounter{#1}
\begin{list}
  {{\alph{#1})}}   
  {\usecounter{#1}\setlength{\rightmargin}{0cm}
   \itemsep 1ex\parsep 1ex}}
\newcommand\bnumlist[1]{\newcounter{#1}
\begin{list}
  {{\bf \arabic{#1})}}   
  {\usecounter{#1}\setlength{\rightmargin}{0cm}
   \itemsep 1ex\parsep 1ex}}
\newcommand\inumlist[1]{\newcounter{#1}
	\begin{list}
		{{\it \arabic{#1})}}   
		{\usecounter{#1}\setlength{\rightmargin}{0cm}
			\itemsep 1ex\parsep 1ex}}
\newcommand\numlist[1]{\newcounter{#1}
\begin{list}
  {{\arabic{#1})}}   
  {\usecounter{#1}\setlength{\rightmargin}{0cm}
   \itemsep 1ex\parsep 1ex}}
\newcommand\romlist[1]{\newcounter{#1}
\begin{list}
  {{(\roman{#1})}}   
  {\usecounter{#1}\setlength{\rightmargin}{0cm}
   \itemsep 1ex\parsep 1ex}}
\newcommand\bromlist[1]{\newcounter{#1}
\begin{list}
  {{\bf \roman{#1})}}   
  {\usecounter{#1}\setlength{\rightmargin}{0cm}
   \itemsep 1ex\parsep 1ex}}
\newcommand\problist{\newcounter{probc}
\begin{list}
  {{\bf \arabic{probc}.}}   
  {\usecounter{probc}\setlength{\rightmargin}{0cm}
   \itemsep 1ex\parsep 1ex}}
\newcommand\problistn[1]{\newcounter{#1}
\begin{list}
  {{\bf \arabic{#1}.}}   
  {\usecounter{#1}\setlength{\rightmargin}{0cm}
   \itemsep 1ex\parsep 1ex}}
\newcommand\subproblist[1]{\newcounter{#1}
\begin{list}
  {{\bf \arabic{probc}.\alph{#1})}}   
  {\usecounter{#1}\setlength{\rightmargin}{0cm}
   \itemsep 1ex\parsep 1ex}}

\usepackage{palatino}
\usepackage{mathpazo}
\usepackage{etoolbox,lipsum}
\usepackage{floatrow}
\usepackage{titlesec}
\titleformat{\chapter}[hang]{\bfseries\Large}{\thechapter}{20pt}{\Large}{}  
\titleformat{\section}[hang]{\bfseries\large}{\thesection}{20pt}{\large}{}
\usepackage{appendix} 
\graphicspath{{./figures/}{Intro/figures/}{Canonical/figures/}{Superposition/figures/}{Linear-forward/figures/}{Linear-inverse/figures/}{Fidelity/figures/}{Statistical/figures/}{Parametric/figures/}{Latent/figures/}{Compressive/figures/}{Machine-Learning/figures/}{Tomography/figures/}{QuantPhase/figures/}{Linear-algebra/figures/}{Fourier/figures/}{Prob-Stats/figures/}{Corr-Funs/figures/}{Optimization/figures/}{GeomOpt/figures/}{WaveOpt/figures/}}
\renewcommand{\vcr}[1]{\bm{\mathrm {#1}}}          
\renewcommand{\psii}{\sst{\psi}{in}}

\renewcommand{\psii}{\sst{\psi}{i}}

\newcommand{\psir}{\sst{\psi}{r}}

\newcommand{\psiG}{\sst{\psi}{G}}

\renewcommand{\Re}[1]{\text{Re} \left\{ \, #1 \, \right\}}

\newcommand{\psid}{\sst{\psi}{d}}

\newcommand{\chii}{\sst{\chi}{i}}
\newcommand{\chid}{\sst{\chi}{d}}
\newcommand{\chis}{\sst{\chi}{s}}
\newcommand{\chit}{\sst{\chi}{t}}
\newcommand{\chiv}{\sst{\chi}{v}}
\newcommand{\phis}{\sst{\phi}{s}}

\newcommand{\tilpsi}{\tilde{\psi}}

\title{Sensitivity fields and parameter estimation \\ from dielectric objects}
\author{George Barbastathis,$^{\dagger,\ddagger}$\\
$\breathe{0}{3} {}^\dagger$Department of Mechanical Engineering  \\
Massachusetts Institute of Technology \\
$\breathe{0}{3} {}^\ddagger$Singapore-MIT Alliance for Research \& Technology\\
Massachusetts Institute of Technology}

\date{June 30, 2024}                                           

\begin{document}
\maketitle

\section*{Abstract}
The quantitative phase image formation process is posed as a problem of parameter estimation from intensity measurements. This approach is inclusive of traditional pixel-oriented imaging, where the sought parameters are the pixel values. The resulting optimization process to find the parameters is then seen to depend on the Sensitivity Field: this is the gradient of the scattered field with respect to the parameters, and it turns out to obey a scattering relationship that is analogous to that of the original scattered field. Examples are given from several regimes of scattering strength. 

\section{Introduction \label{sec:par-est-ph-obj-intro} }

``Quantitative Phase Imaging'' has become, not least thanks to the tireless work of our late colleague and friend Gabi Popescu\ \cite{inv:quant-ph-img-2018}, whose memory this special issue honors, a standard and popular sub-discipline of Imaging Science. It is the topic of dedicated technical meetings and special sessions within bigger conferences, and of course numerous theses and government grants. Taken at face value, the topical term suggests as its ultimate goal the numerically accurate reconstruction of the spatial details of electromagnetic field's phase. 

The question of phase, and of its meaning and utility for imaging, dates back to at least eighty years ago. This was when the origins of present optical, x-ray and electron microscopy methods were being set by Lawrence Bragg and others. In the words of Denis Gabor\ \cite{inv:Gabor49micro-reconstr}:
\begin{quote}
	``[...] Bragg's method, in which a lattice is reconstructed by diffraction from an X-ray diffraction pattern, can be applied only to a rather exceptional class of periodic structures. It is customary to explain this by saying that diffraction diagrams contain information on the intensities only, but not on the phases. The formulation is somewhat unlucky, as it suggests at once that since phases are unobservables, this state of affairs must be accepted. In fact, not only that part of the phase which is unobservable drops out of conventional diffraction patterns, but also the part which corresponds to geometrical and optical properties of the object, and which in principle could be determined by comparison with a standard reference wave.''
\end{quote}
Already in this introductory paragraph, and many times subsequently in this remarkable paper, Gabor makes the point that the phase is not a goal in its own right, but rather as the means to recover more information about the {\em object} of interest. 

Of course, one important difference between Gabor's time and ours is that then image interpretation was done entirely manually; therefore, it was reassuring for the human processors, scientists and their assistants alike, to be operating on visually detailed images whose fidelity to the true scene would be somehow guaranteed. Nowadays, image exploitation is largely automated, and yet we still treat the processing algorithms ``anthropomorphically''; that is, in equal need of spatial richness and accuracy. Even more importantly, the intervening development of Information Theory---with considerable contributions by Gabor himself---has taught us that the information content of an object is limited by its statistical variation. Spatial variability that is irrelevant to a specific information retrieval task is, at best, unnecessarily resource demanding and, in some cases, it may be outright harmful. 

As an example, consider the task of localizing a blob-like object whose blob shape is already known. The anthropomorphic way would be to image the blob as accurately as possible, and then find its center of mass. In the 1960's this would have involved calculators and rulers. Nowadays, with a computer one might first try to ``improve'' the image as much as possible with regularizing filters and the like; and then apply a centroid or similar algorithm. This seems like a lot of work just to get out a single coordinate---the center of mass. Moreover, even though my students and I followed this exact same approach before and showed that at least our specific instance was accurate\ \cite{inv:liu12-subpixel1D,inv:liu14-subpixel2D}, in general there is no guarantee that the image ``improvement algorithm'' is bias-free. 

On the other hand, it is possible to imagine methods of determining the center of mass that are impervious to shape specifics. In fact, one of the earliest ``optical computing'' methods, the van der Lugt correlator\ \cite[sec.~8.4]{book:fourieroptics2}, does just that. Thus, it is evident that a gap exists between the traditional imaging mentality that prioritizes exquisite spatial detail even if only a few parameters are to be extracted from it; and the information theoretic-driven view that acknowledges the sparsity of the sought-after description and orients the retrieval method accordingly. 

My purpose in the present paper is to bridge this gap by recasting spatial imaging as a problem of parameter estimation. This includes fully spatially resolved images as special cases. When the parameters of interest are significantly fewer than the number of pixels, certain efficiencies become possible for imaging system design: it is no accident that the parametric view is popular in modalities deprived of the benefit of high pixel-count spatial reconstructions, e.g. the geosciences.

The core development of the parametric approach is in Section~\ref{sec:par-est-sp-img}. It consists of formulating an optimization functional on the detected intensity---which is always the case, even if the estimate of a phase-like quantity is to be sought after later as an intermediate product. Taking gradients on the functional reveals the need for a Sensitivity Field, which is simply defined as the gradient of the detected field to the sought-after parameters. The Sensitivity Field is an artificial, mathematical construct; however, it turns out to obey propagation rules similar to the physical fields whose gradients it represents. 

Section~\ref{sec:par-est-sensi-mod} derives the Sensitivity Fields for three regimes that are commonly assumed in light-matter interaction: the thin-film approximation, the first Born (or Wolf) kernel, and strong scattering approaches such as the Lippmann-Schwinger equation and the Beam Propagation Method. Interestingly, the Sensitivity Fields themselves scatter similar to and driven by the electromagnetic fields they originated from. These results are analogous, albeit not entirely identical to sensitivity and adjoint analysis of dynamical systems. \cite{book:pontryagin-optimal,dyn:cao03-adjoint-sensi}

Before diving into the technical details, let me note what the paper is {\em not} attempting to do. First, it is not meant to be an exhaustive analysis of electromagnetic scattering models, and it could not credibly be in such a short space. Several excellent textbooks cover this topic in detail\ \cite{book:tatarski1961}. The purpose of using a few such scattering models is to show how their Sensitivity Functions are derived. 

Second, my goal here is not to compare the validity or fidelity of different parametric descriptions or scattering models. The former requires a careful formulation of detection noise statistics, which is beyond the scope of the present paper. The basics of the latter are in the aforementioned textbooks, and still subject to active research. \cite{inv:pang-unified} Lastly, in Section~\ref{sec:par-est-sp-img} I only use a simple quadratic---that is, optimal for additive Gaussian statistics only---loss function to derive the sensitivity. Examining the behavior of the Sensitivity Field with alternative loss functions, e.g. adversarial, and more generally adoption of modern optimization methods would be interesting topics for follow-up work.

\section{Parameter estimation from spatial images \label{sec:par-est-sp-img} }

\subsection{Intensity detection and the sensitivity field \label{sec:par-est-loss-fun} }

\begin{figure}[t]
	\centergraphics{0.77}{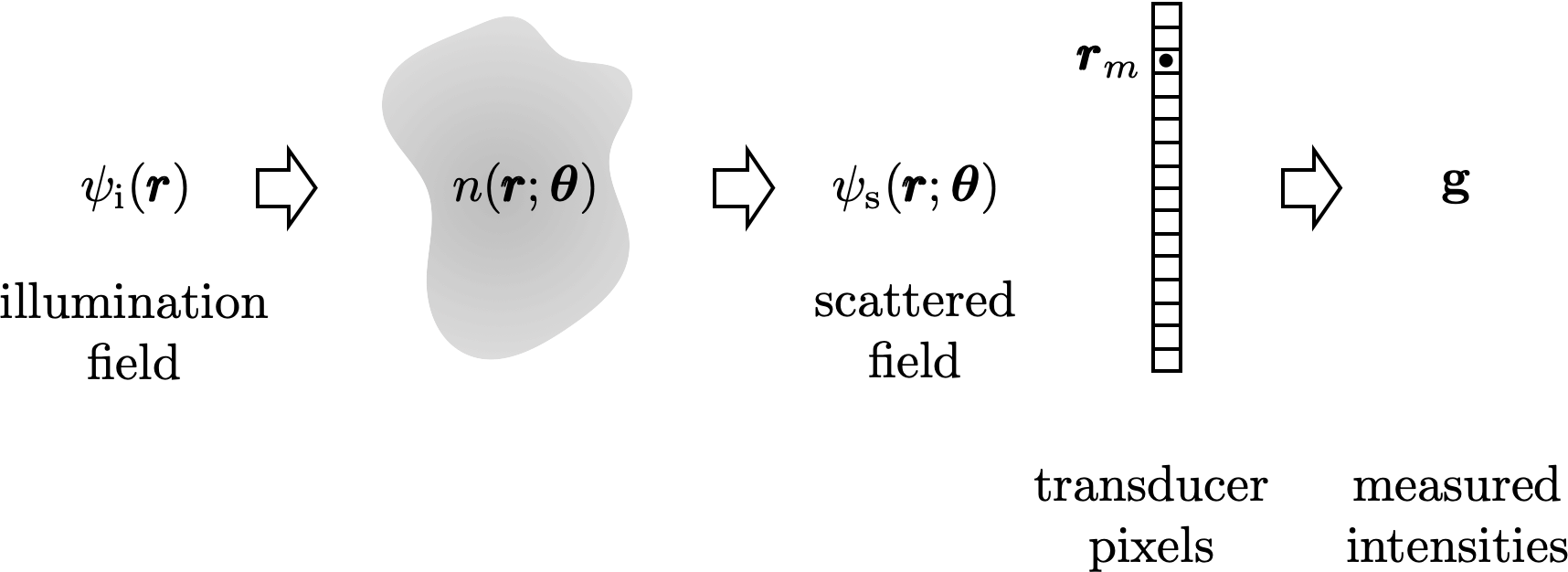}
	\caption{General scattering geometry with a single intensity detection on a photoelectric transducer. \label{fig:scattering-geom} }
\end{figure}

The geometry to be used in the subsequent analysis is shown in Figure~\missing{Make.} Let $\psii(\vcf{r})$ denote the incident electromagnetic field as function of the Cartesian spatial coordinate $\vcf{r}$; and let $n(\vcf{r};\vcrg{\theta})$ denote the refractive index of the object whose parameters $\vcrg{\theta}$ are of interest. The refractive index is assumed to be real, \ie\ signifying a pure dielectric object. The parameters $\vcrg{\theta}$ may be thought of as ``controlling'' the spatial distribution of the refractive index; some examples are given in Section~\ref{sec:par-est-para-exs} below. 

The resulting scattered field is denoted as $\psis(\vcf{r};\vcrg{\theta})$ and transduced at $M$ pixel locations $\vcf{r}_m,$ $m=1,\ldots,M,$ where the intensities 
\begin{equation}
	g_m(\vcrg{\theta}) = \left|\psis(\vcf{r}_m;\vcrg{\theta})\right|^2
	\label{eq:par-est-g-m-def}
\end{equation}
are recorded. The vector
\begin{equation}
	\vcr{g}(\vcrg{\theta}) = \left( \breathe{0.5}{3} 
	\begin{array}{ccccc} g_1(\vcrg{\theta}) & \cdots & g_m(\vcrg{\theta}) & \cdots & g_M(\vcrg{\theta})
		\end{array}  \right)^\dagger 
\end{equation}
is ``the ideal measurement,'' where $\dagger$ denotes the transpose. The right-hand side in (\ref{eq:par-est-g-m-def}) is the the physical model for light propagation through the object for {\em any} choice of parameters $\vcrg{\theta}$. It is also known as the ``forward operator.'' In optical systems usually the $\vcf{r}_m$'s are arranged on a planar 2D lattice of pixels, but in other modalities, e.g. in geo-imaging, they may be randomly placed. 

In a noiseless measurement and with perfect knowledge of $\vcrg{\theta}$, eq.~\ref{eq:par-est-g-m-def} holds with strict equality. In practice, a noisy measurement $\vcr{\tilde{g}}$ will be instead recorded; and the parameter vector $\vcrg{\theta}$ is to be determined. To carry out this latter task, define a loss function such as
\begin{equation}
	E(\vcrg{\theta}) = \frac{1}{4M} \: \sum_{m=1}^{M} 
	\left[ \breathe{0.5}{2.5} \tilde{g}_m - \left|\psis(\vcf{r}_m;\vcrg{\theta})\right|^2 \right]^2.
	\label{eq:par-est-loss-fun-def}
\end{equation}
It is useful to reiterate that inside the square bracket the first term is the noisy measurement, whereas the second term is the forward model, parameterized according to $\vcrg{\theta}$. The quadratic loss function is optimal in the maximum likelihood sense for additive Gaussian detection statistics. Other classical choices, e.g. for Poisson (shot-noise) statistics, or more modern ones such as adversarial loss are possible, but I will not work them out in detail in this paper. 

The optimal parameter vector is found from solving the minimization problem
\begin{equation}
	\vcrg{\hat{\theta}} = \: \argmin{\vcrg{\theta}} E(\vcrg{\theta}). 
	\label{eq:par-est-loss-fun-mini}
\end{equation}
If additional prior information on the parameters can be captured in the form or a regularizer $\Phi(\vcrg{\theta})$, then the minimization problem is modified as the unconstrained  
\begin{equation}
	\vcrg{\hat{\theta}} = \: \argmin{\vcrg{\theta}} \left\{ \breathe{0.5}{2.5} E(\vcrg{\theta}) + \kappa \: \Phi(\vcrg{\theta}) \right\}  ; 
	\label{eq:par-est-loss-fun-mini-reg}
\end{equation}
or the inequality-constrained 
\begin{equation}
	\vcrg{\hat{\theta}} = \: \argmin{\vcrg{\theta}} \Phi(\vcrg{\theta}) \quad 
	\text{subject to} \quad E(\vcrg{\theta}) < \kappa.
	\label{eq:par-est-loss-fun-mini-reg-unconstr}
\end{equation}
In either option, $\kappa$ is acting as a regularization parameter. In the interest of brevity, the regularized forms (\ref{eq:par-est-loss-fun-mini-reg}-\ref{eq:par-est-loss-fun-mini-reg-unconstr}) are largely left unexploited in this paper. 

Existence and uniqueness of minima need to be analyzed on a case-by-case basis, depending on the forward problem and its parameterization. If a minimum does exist, it may be found in a number of ways: e.g. genetic algorithms, gradient-based methods, etc. The simplest among the latter, the gradient descent, proceeds iteratively as
\begin{equation}
	\vcrg{\hat{\theta}}[t+1] = \vcrg{\hat{\theta}}[t] - \eta[t] \, \pder{E\left(\vcrg{\hat{\theta}}[t]\right)}{\vcrg{\theta}}.
\end{equation}
Here, $t$ denotes the iteration index and $\eta$ is the learning rate. The gradient term is obtained from (\ref{eq:par-est-loss-fun-def}) as
\begin{equation}
	\pder{E\left(\vcrg{\hat{\theta}}[t]\right)}{\vcrg{\theta}} = \frac{1}{M} \sum_{m=1}^M 
	\Re{\psis^*(\vcf{r}_m) \: \pder{\psis(\vcf{r}_m)}{\vcrg{\theta}} } \: 
	\left[ \breathe{0.5}{2.5} \tilde{g}_m - \left|\psis(\vcf{r}_m;\vcrg{\theta})\right|^2 \right].
	\label{eq:par-est-loss-fun-deri}
\end{equation}
It is evidently convenient at this point to introduce the Sensitivity Field 
\begin{equation}
	\chis(\vcf{r};\vcrg{\theta}) \equiv \pder{\psis(\vcf{r};\vcrg{\theta})}{\vcrg{\theta}},
	\label{eq;par-est-sensi-fun-def}
\end{equation}
which is simply the gradient of the scattered field with respect to the chosen parameterization; and the local error term
\begin{equation}
	\Delta_m(\vcrg{\theta}) \equiv  \tilde{g}_m - \left|\psis(\vcf{r}_m;\vcrg{\theta})\right|^2.
	\label{eq:par-est-local-error-def}
\end{equation}
In terms of these two newly defined quantities, the loss function gradient is written more compactly as 
\begin{equation}
	\pder{E}{\vcrg{\theta}} = \frac{1}{M} \sum_{m=1}^M 
	\Re{\breathe{0.5}{2.5} \psis^*(\vcf{r}_m;\vcrg{\theta}) \: \chis(\vcf{r}_m;\vcrg{\theta}) } \: \Delta_m(\vcrg{\theta}).
	\label{eq:par-est-loss-fun-sensi}
\end{equation}

\subsection{Multiple exposures \label{sec:par-est-multi} }

\begin{figure}[t]
	\centergraphics{0.77}{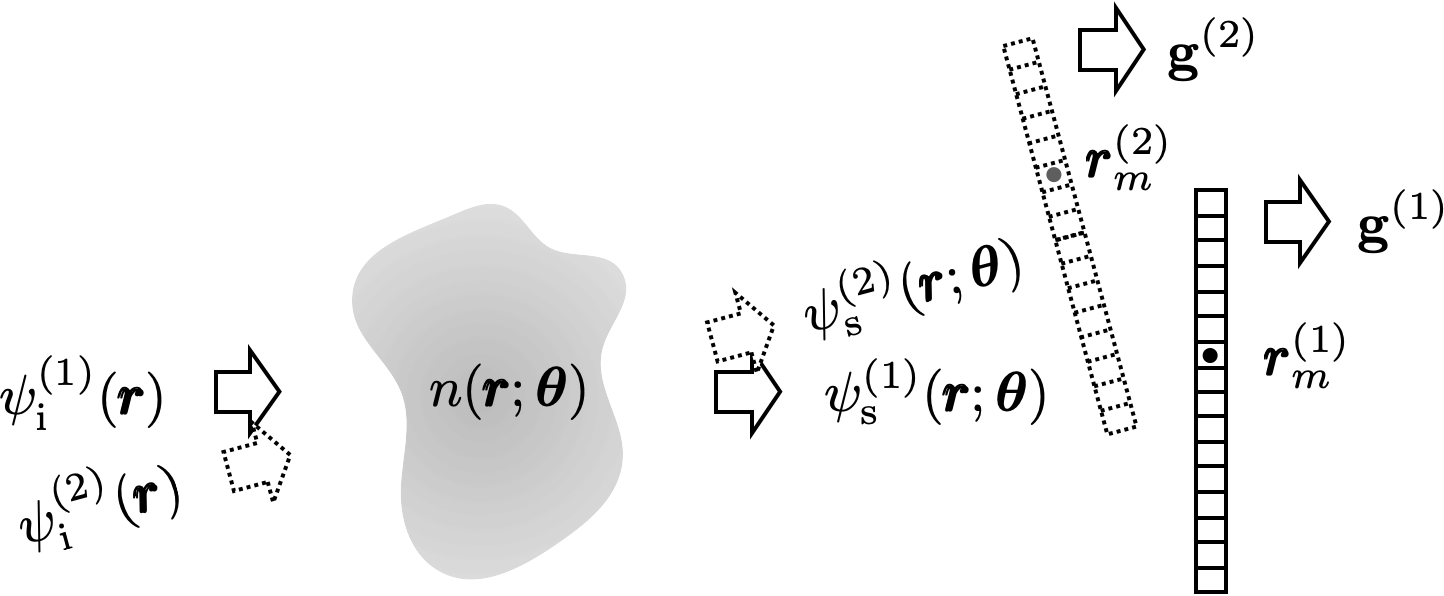}
	\caption{General scattering geometry with several intensity detection steps, each time with the illumination and photoelectric transducer at a different orientation or rotation. \label{fig:scattering-geom-multi} }
\end{figure}

For the sake of completeness, I develop this notation a little bit further for the numerous methods that employ multiple exposures onto the same object. Popular examples are confocal microscopy, transport of intensity, ptychography, computed (Radon) tomography, optical diffraction tomography, optical coherence tomography, etc. A drawing suggestive of rotations between exposures, as in tomography, is shown in Figure~\ref{fig:scattering-geom-multi}. Let $l$ denote the counter of exposures, entered as superscript in the other notations. Thus, $\vcf{r}_m^{(l)}$ is the coordinate of the $m$-th detector pixel at the $l$-th exposure, $\psii^{(l)}(\vcf{r})$ is the illuminatioun at the $l$-th exposure, etc. The loss function is then expressed as
\begin{equation}
	E(\vcrg{\theta}) = \frac{1}{4LM} \sum_{l=1}^L w^2_l \sum_{m=1}^M 
	\left[ \breathe{0.5}{2.5} \tilde{g}^{(l)}_m - 
	\left| \psis^{(l)}(\vcf{r}^{(l)}_m;\vcrg{\theta}) \right|^2 \right]^2.
	\label{eq:par-est-loss-fun-multi}
\end{equation}
Here, for full generality we have weighed different exposures with the positive numbers $w^2_l$. These express relative confidence, for example noisier exposures might receive lower weights. 

All the derivations from the previous section follow similarly, finally obtaining 
\begin{equation}
	\pder{E}{\vcrg{\theta}} = \frac{1}{LM} \sum_{l=1}^L w^2_l \sum_{m=1}^M 
	\Re{\breathe{0.5}{2.5} \psis^{(l)\,*}(\vcf{r}^{(l)}_m;\vcrg{\theta}) \: \chis^{(l)}(\vcf{r}^{(l)}_m;\vcrg{\theta}) } \: \Delta^{(l)}_m(\vcrg{\theta}).
	\label{eq:par-est-loss-fun-sensi-multi}
\end{equation}
This formulation is appealing because it is compatible with tensorial formulations in computing platforms such as Python and Julia. 

\subsection{Holographic detection and the sensitivity field 
	\label{sec:par-est-holo-det} }
	
\begin{figure}[t]
	\centergraphics{0.77}{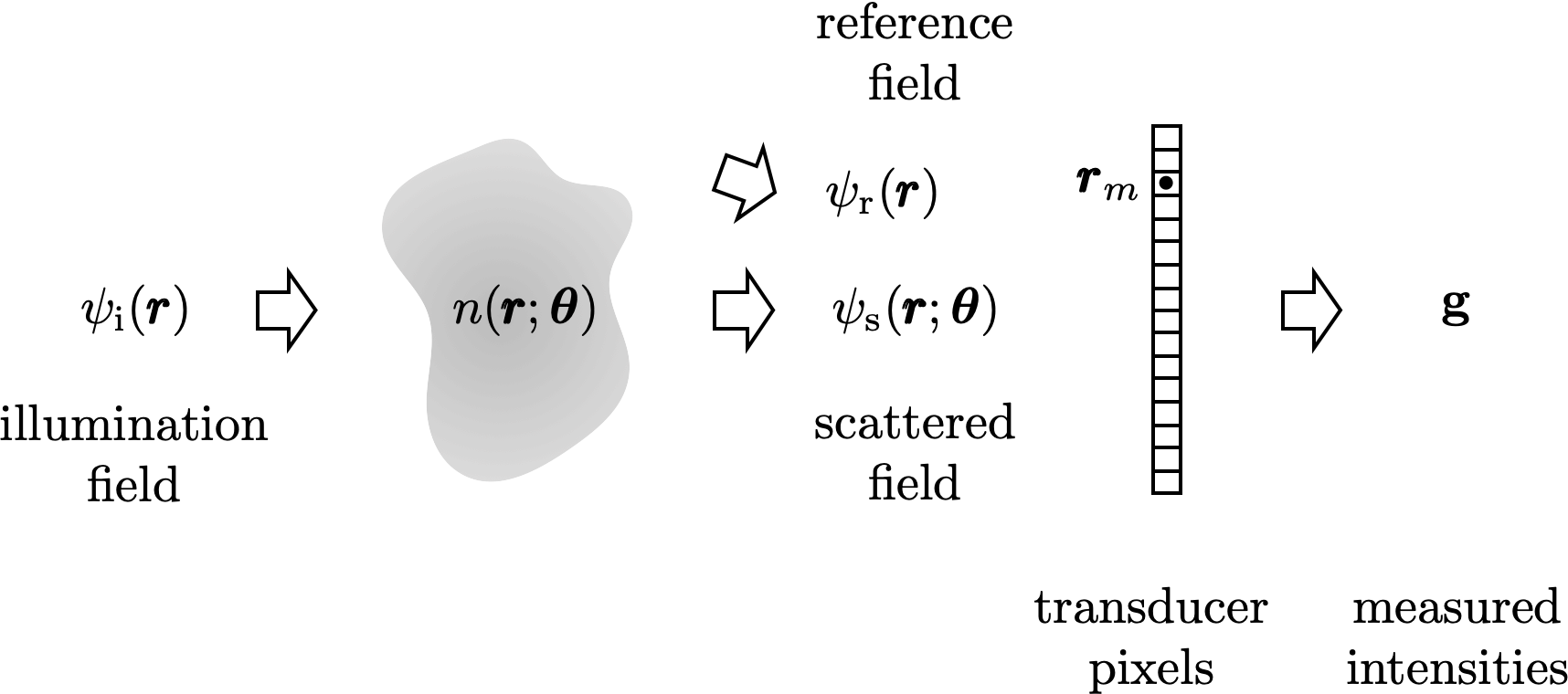}
	\caption{General scattering geometry with holographic detection. \label{fig:scattering-geom-holo} }
\end{figure}

In the holographic configuration, rather than detecting the intensity scattered field itself, one interferes it with a reference field $\psir(\vcf{r})$. Thus, the intensity at the $m$-th pixel is the interferogram
\begin{equation}
	g_m = \left| \psir(\vcf{r}) + \psis(\vcf{r};\vcrg{\theta})  \right|^2
	\label{eq:par-est-g-m-holo}
\end{equation}
It is also common in experiments for the reference power to be much stronger than the scattered field; that is, $|\psir|\gg|\psis|$ everywhere. We then obtain
\begin{equation}
	\pder{E}{\vcrg{\theta}} \approx \frac{1}{M} \sum_{m=1}^M 
	\Re{\breathe{0.5}{2.5} \psir^*(\vcf{r}_m;\vcrg{\theta}) \: \chis(\vcf{r}_m;\vcrg{\theta}) } \: \Delta_m(\vcrg{\theta}).
	\label{eq:par-est-loss-fun-sensi-holo}
\end{equation}

Whether (\ref{eq:par-est-loss-fun-sensi-holo}) is an improvement over the reference-free version (\ref{eq:par-est-loss-fun-sensi}) depends on the nature of the scattered field. For slowly-varying fields, the reference term acts as an amplification factor on the sensitivity, which should be beneficial. On the other hand, the high frequencies in the scattered field would make the interference vary too rapidly, and that might lead to local minima in the loss function. A detailed comparison of direct {\it vs.} holographic intensity detection is beyond the scope of this paper, but certainly very interesting for future work.

\subsection{Partially coherent illumination \label{sec:par-est-part-coh} }

So far the illumination has been assumed to be temporally and spatially coherent. To account for spatial partial coherence, one should replace the field in (\ref{eq:par-est-g-m-def}) with 
\begin{equation}
	g_m(\vcrg{\theta}) = \sst{\Gamma}{I}(\vcf{r}_m,\vcf{r}_m;\vcrg{\theta}).
	\label{eq:par-est-g-m-def-part-coh}
\end{equation}
Here, $\sst{\Gamma}{I}(\vcf{r}_1,\vcf{r}_2)$ is the spatial correlation function of the scattered field, more usually referred to as mutual intensity, at a general pair of spatial coordinates $\vcf{r}_1$, $\vcf{r}_2$. This may also be expressed in terms of the coherent modes \missing{Ref} as
\begin{equation}
	\sst{\Gamma}{I}(\vcf{r}_1,\vcf{r}_2;\vcrg{\theta}) = 
	\sum_{q=1}^\infty \beta_q(\vcrg{\theta}) \: \zeta_q(\vcf{r}_1;\vcrg{\theta}) \: \zeta^*_q(\vcf{r}_2;\vcrg{\theta}). 
\end{equation}
The ideal measurement then is
\begin{equation}
	g_m(\vcrg{\theta}) = \sum_{q=1}^\infty \beta_q(\vcrg{\theta}) \left|\zeta_q(\vcf{r}_m;\vcrg{\theta})\right|^2.
	\label{eq:par-est-g-m-def-coh-modes}
\end{equation}
Expression (\ref{eq:par-est-g-m-def-part-coh}) or (\ref{eq:par-est-g-m-def-coh-modes}) is then to be substituted in the loss function (\ref{eq:par-est-loss-fun-def}) and subsequent formulae. In the interest of keeping the present paper confined, a full analysis of parameter estimation under partially coherent illumination is left for future work.

\subsection{On the choice of parameterization \label{sec:par-est-para-exs} }

The following three examples elucidate how the parameterization choice influences the formulation of the estimation problem.

\bnumlist{sensi-mods}
\item {\bf Traditional pixel-oriented spatial imaging.} Conceding that at best one may only hope to recover a bandlimited version of the refractive index, let 
\[
\left\{ \vcf{r}_j \right\}_{j=1,\ldots,J}
\]
denote a sampling lattice at the object space, and 
\begin{equation}
	n(\vcf{r}) = \sum_{j=1}^{J} f_j \: \delta(\vcf{r} - \vcr{r}_j).
	\label{eq:par-est-sampled-def}
\end{equation}
The coefficients $f_j$ are what traditionally one might call the spatial pixel values of the (quantitative) phase in 2D, or of the refractive index reconstruction in 3D. By assembling these values into the parameter vector
\[
\vcrg{\theta} = \left( \breathe{0.5}{3} 
\begin{array}{ccccc} f_1 & \cdots & f_j & \cdots & f_J
\end{array}  \right)^\dagger, 
\]
one can see that the minimum of (\ref{eq:par-est-loss-fun-def}), if it exists, will correspond to the traditional spatial ``quantitative phase image.''

\begin{figure}[t]
	\centergraphics{0.29}{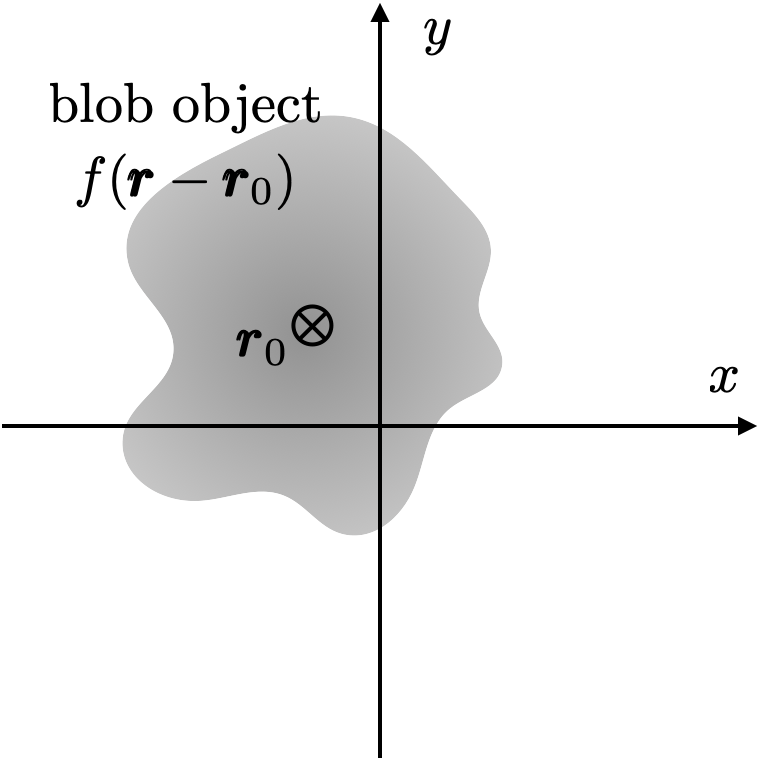}
	\caption{Geometry for estimating the displacement $\vcf{r}_0$ of a known blob shape $f(\vcf{r})$. \label{fig:blob-displ} }
	\end{figure}
	
Note that if $LM>J$, then the problem is overdetermined, \ie\ one has more measurements than unknowns; whereas, if $LM<J$ then it is ill-posed, or underdetermined. Alternatively, (\ref{eq:par-est-loss-fun-def}-\ref{eq:par-est-loss-fun-mini}) may be thought of as solving a set of $M$ nonlinear equations with $J$ unknowns.
\item {\bf Localization of a known shape.} Returning to the localization example from the Introduction section, suppose that 
\begin{equation}
	n(\vcf{r};\vcf{r}_0) = f(\vcr{r}-\vcf{r}_0),
\end{equation}
where $f(\vcf{r})$ is a known shape, and it is desired to localize it; \ie, the only unknown is the displacement $\vcf{r}_0$. This geometry is shown in Figure~\ref{fig:blob-displ}. The parameterization with respect to displacement is
\[
\vcrg{\theta} = \vcf{r}_0. 
\]
If, moreover, the imaging system is spatially shift-invariant, then (\ref{eq:par-est-loss-fun-sensi}) reduces to
\begin{equation}
	\pder{E}{\vcrg{\theta}} = \frac{1}{2M} \sum_{m=1}^M 
	\pder{\left|\psis(\vcf{r}-\vcrg{\theta})\right|^2}{\vcf{r}}  \: \Delta_m(\vcrg{\theta}).
	\label{eq:par-est-centro-sensi}
\end{equation}
The problem has become one of finding the best fit displacement coordinate vector to the forward model for the $LM$ intensity measurements. As long as $LM>3$ (or $2$ for planar blobs) this estimation problem is overdetermined. 
\item {\bf Spatial representation in terms of basis functions.} Compressive Sensing relies on the existence of a set of basis functions $c_j(\vcf{r})$ where the object can be represented efficiently. For the refractive index as object, this is expressed as
\begin{equation}
	n(\vcf{r}) = \sum_{j=1}^{J} f_j \, c_j(\vcf{r})
	\label{eq:par-est-basis-def}
\end{equation}
where most of the coefficients $f_j$ should turn out to be zero. Such sparse bases are discovered by unsupervised learning, e.g. dictionaries\ \missing{Ref} The pixel-sampled parameterization (\ref{eq:par-est-sampled-def}) is seen to be a special albeit not necessarily sparse version of (\ref{eq:par-est-basis-def}). The parameter vector is similarly chosen as 
\[
\vcrg{\theta} = \left( \breathe{0.5}{3} 
\begin{array}{ccccc} f_1 & \cdots & f_j & \cdots & f_J
\end{array}  \right)^\dagger. 
\]
Now a regularized approach such as (\ref{eq:par-est-loss-fun-mini-reg}) or (\ref{eq:par-est-loss-fun-mini-reg-unconstr}) suggests itself, with a suitable choice of sparsity-promoting regularizer, e.g. $\Phi(\vcrg{\theta}) = \lonenorm{\vcrg{\theta}}.$
\end{list}

\section{Sensitivity Fields for some scattering models \label{sec:par-est-sensi-mod} }

The Sensitivity Field (\ref{eq;par-est-sensi-fun-def}) in general may be obtained numerically by automatic differentiation. Instead, it is worthwhile to derive explicit expressions for a  few common cases of scattering strengths: the thin-film object, the single-event scatterer or First Born approximation, and the multiple scattering regime. The derivations are simple and yield results with a certain intuitive appeal. Moreover, they may provide insights about convergence and fidelity in future real-life applications.  

The starting point is the spatially modulated scalar Helmholtz wave equation
\begin{equation}
	\nabla^2 \psi(\vcf{r}; \vcrg{\theta}) + 
	k^2 \, n^2(\vcf{r}; \vcrg{\theta}) \, \psi(\vcf{r}; \vcrg{\theta}) = 0, 
	\label{eq:sca-wave-eq}
\end{equation}
where $k$ is the free-space wavenumber and $\psi(\vcf{r})$ the appropriate element of the electric field vector.  Eq.~\ref{eq:sca-wave-eq} already neglects polarization effects due to propagation at large numerical apertures and optical anisotropy. 


\subsection{Point-spread functions \label{sec:par-est-psfs} }

Before developing the Sensitivity Fields, one important detail needs to be taken care of. In the following, $\psiG(\vcf{r})$ denotes the optical system's Green function, also known as impulse response and point-spread function (PSF). Some typical Green's functions are listed below and may be replaced in any of the scattering formulae of subsequent Sections, as appropriate. 
\begin{itemize}
	\item The free-space Green's function
	\begin{equation}
		\psiG(\vcf{r}) = \frac{\expb{ik\left|\vcf{r}\right|}}{ik\left|\vcf{r}\right|},
		\label{eq:par-est-sphwa-kern}
	\end{equation}
which is an ideal spherical wavefront. Since an ideal point source does not exist, this expression does not represent a physically realizable solution to the Helmholtz equation (\ref{eq:sca-wave-eq}) for electromagnetic fields; closest to a realistic solution is the dipole field\ \missing{Ref.} Nevertheless, for numerical apertures less than $60^\circ$ the ideal spherical wave approximation works quite well. 
\item The paraxial free-space Green's function, also referred to as the Fresnel kernel
	\begin{equation}
		\psiG(\vcf{r}) = \expb{ik \, z + ik \, \frac{x^2+y^2}{2z}}.
		\label{eq:par-est-fresnel-kern}
	\end{equation}
\item The ideal in-focus diffraction-limited system PSF
	\begin{equation}
		\psiG(x,y) = \text{jinc}\left( \frac{\sqrt{x^2+y^2}}{\sst{R}{DL}} \right),
	\end{equation}
	where $\sst{R}{DL}$ is the radius of the diffraction-limited spot and $\text{jinc}(\cdot)$ is the same function as in \missing{Ref specific equation in Goodman}.
\end{itemize}

\subsection{``Phase objects'' and the thin-film approximation \label{sec:par-est-ph-obj-what} }

\begin{figure}[t]
	\centergraphics{0.47}{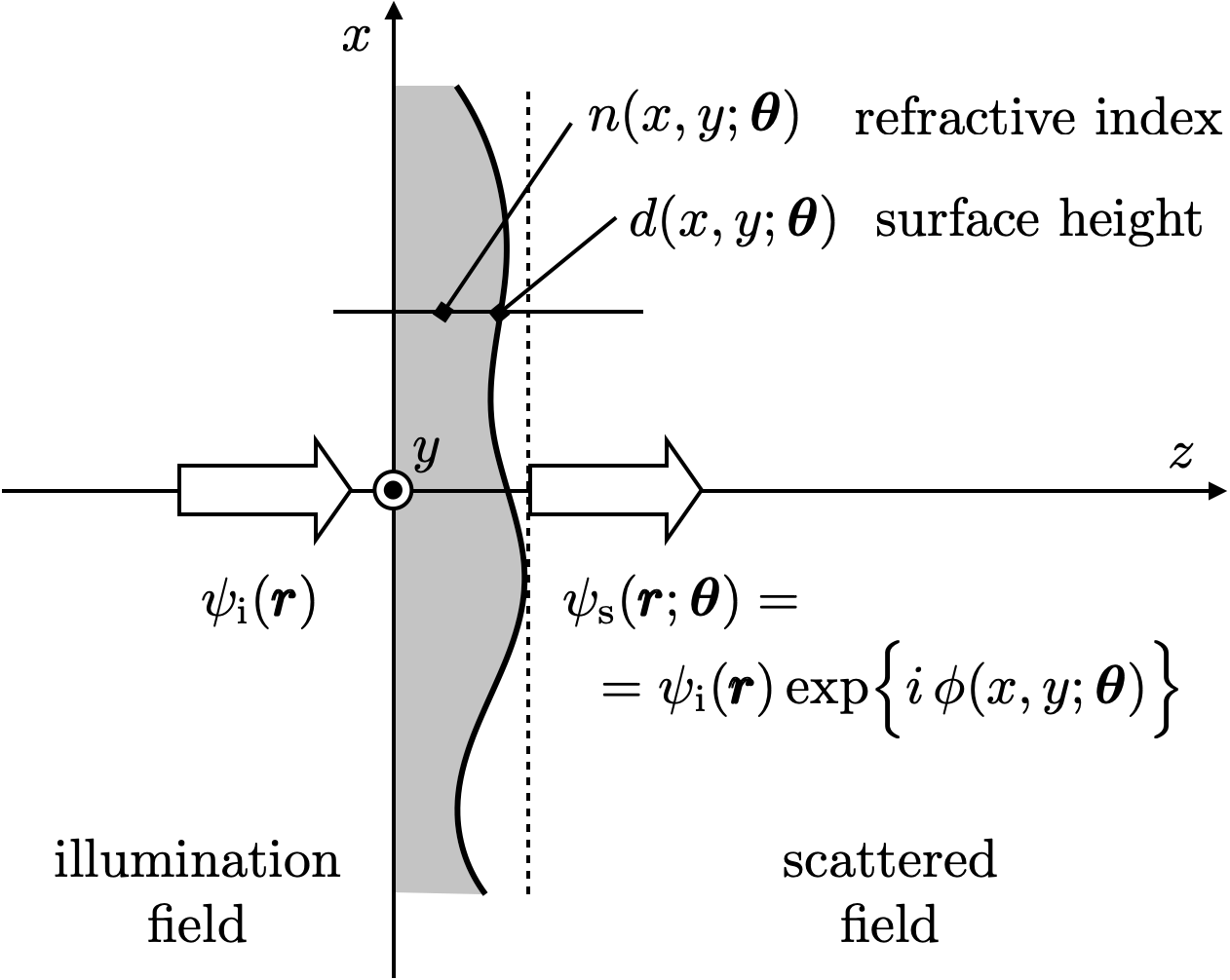}
	\caption{The thin transparency is an object that imparts pure phase modulation onto an incident field. \label{fig:thin-trans-phi} }
\end{figure}

Loosely speaking, a phase object is generally understood as a spatial material distribution that interacts with light in such a way as to impart phase delay only on the wavefield; no attenuation is to be imparted anywhere. A good approximation for such an object is an infinitessimally thin film yet with a thickness along the optical axis that is modulated as $d(x,y;\vcrg{\theta})$ as function of the lateral spatial coordinates $x, y$; and whose index of refraction $n(x,y;\vcrg{\theta})$ is purely real, and also modulated. This configuration is depicted in Figure~\ref{fig:thin-trans-phi}. 

It is customary to define the optical path difference (OPD) 
\begin{equation}
	\phi(x,y;\vcrg{\theta}) = k \, \left[ \breathe{0.5}{2.5} n(x,y;\vcrg{\theta})  - 1 \right]
	\, d(x,y;\vcrg{\theta})
	\label{eq:par-est-opd}
\end{equation}
and say that the wavefront itself is modulated along the lateral dimensions $(x,y)$ according to the OPD. That is, the field emanated directly after the rightmost edge of the thin transparency, at the vertical dashed line in Figure~\ref{fig:thin-trans-phi}, is
\begin{equation}
	\psis(\vcf{r};\vcrg{\theta}) = \psii(\vcf{r}) \, \expb{\breathe{0.5}{2.5} i\,\phi(x,y;\vcrg{\theta}) }.
	\label{eq:par-est-thin-film-psis} 
\end{equation}
A brief proof of (\ref{eq:par-est-thin-film-psis}) including the list of conditions that have to be met for it to be acceptable is in the Appendix. 

Past the sample, the field propagates either in free space or through an optical system till it reaches the detector. In both cases, we denote the point-spread function as $\psiG(\vcf{r})$, as discussed in Section~\ref{sec:par-est-psfs}. The final field on the detector is
\begin{equation}
	\psid(\vcf{r};\vcrg{\theta}) = \iint \psii(x',y',z) \, \expb{\breathe{0.5}{2.5} i\,\phi(x,y;\vcrg{\theta}) } \psiG(x-x',y-y') \: \dx'\dy'.
	\label{eq:par-est-thin-film-psid}
\end{equation}
Note that for applications utilizing the thin film approach, such as imaging of biological cells, it may be necessary to parameterize both refractive index and thickness. This is even though they cannot be decoupled from a single intensity measurement and in the absence of other priors, as is evident from (\ref{eq:par-est-opd}).

Leaving this issue aside, the Sensitivity Field is readily obtained as
\begin{equation}
	\chid(\vcf{r}; \vcrg{\theta}) = \iint \chii(x',y',z; \vcrg{\theta}) \, \expb{\breathe{0.5}{2.5} i\,\phi(x,y;\vcrg{\theta}) } \psiG(x-x',y-y') \: \dx'\dy',
	\label{eq:par-est-thin-film-sensi-illum}
\end{equation}
where the ``Sensivity illumination'' field is
\begin{align}
	\chii(\vcf{r}; \vcrg{\theta}) & = \psii(\vcf{r}) \: \pder{\phi(x,y;\vcrg{\theta})}{\vcrg{\theta}}
		\label{eq:par-est-thin-film-chii-phi} \\
		& = \psii(\vcf{r}) \: \left\{ \breathe{1}{3} \pder{n(x,y;\vcrg{\theta})}{\vcrg{\theta}} \, d(x,y;\vcrg{\theta}) + \left[ \breathe{0.5}{2.5} 
		n(x,y;\vcrg{\theta})-1\right]  \pder{d(x,y;\vcrg{\theta})}{\vcrg{\theta}} \right\}.
		\label{eq:par-est-thin-film-chii-n-d}
\end{align}
Evidently, the Sensitivity Field is the same as the field produced by the same object when the illumination is modulated by the OPD sensitivity $\partial \phi/\partial \vcrg{\theta}$. This interesting symmetry will recur in Section~\ref{sec:par-est-first-Born} with the first Rytov approximation (\ref{eq:par-est-Rytov-sensi}) and in Section~\ref{sec:par-est-lipp-schw} with the Lippmann-Schwinger integral equation (\ref{eq:par-est-lipp-schw-sensi}).

Alternatively, the terms within the integral may be regrouped as
\begin{equation}
	\chid(\vcf{r}; \vcrg{\theta}) = \iint \psii(x',y',z; \vcrg{\theta}) \, \expb{\breathe{0.5}{2.5} i\,\chit(x,y;\vcrg{\theta}) } \psiG(x-x',y-y') \: \dx'\dy',
	\label{eq:par-est-thin-film-sensi-transp}
\end{equation}
where now the phase transmissivity is
\begin{equation}
	\chit(x,y;\vcrg{\theta}) = \phi(x,y;\vcrg{\theta}) - i\log \left(\pder{\phi(x,y;\vcrg{\theta})}{\vcrg{\theta}}\right).
\end{equation}
Due to the imaginary part in this expression, this is no longer a pure ``phase object;'' it includes an amplitude modulation. An analogous result occurs in Section~\ref{sec:par-est-first-Born} with the first Born approximation (\ref{eq:par-est-first-Born-sensi}).

\subsection{First Born and First Rytov approximations \label{sec:par-est-first-Born} }

The Born series is a rigorous, if not necessarily stable numerically method to obtain a solution for the scattered field in (\ref{eq:sca-wave-eq}). Keeping the first term only implies that multiple scattering is negligible. This results from the substitution 
\begin{equation}
	\psi(\vcf{r}; \vcrg{\theta}) = \psii(\vcf{r}) + \psis(\vcf{r}; \vcrg{\theta})
\end{equation}
for the total field, and the approximation 
\begin{equation}
	\psis(\vcf{r};\vcrg{\theta}) \approx \int \psii(\vcf{r}) \, v(\vcf{r'};\vcrg{\theta}) \, 
	\sst{\psi}{G}(\vcf{r}-\vcf{r'}) \, \text{d}\vcf{r'}.
	\label{eq:par-est-first-Born}
\end{equation}
Here, 
\begin{equation}
	v(\vcf{r};\vcrg{\theta}) \equiv \frac{1}{2} \left[ n^2(\vcf{r};\vcrg{\theta}) - 1 \right]
	\label{eq:par-est-scattering-potential}
\end{equation}
is the scattering potential. Result (\ref{eq:par-est-first-Born}) is known as the First Born approximation or  Born integral and more recently it has started being referred to as the Wolf transform. \missing{Ref Wolf 1967} 

It is straightforward to obtain the Sensitivity Field for the First Born approximation as 
\begin{equation}
	\chis(\vcf{r}; \vcrg{\theta}) = \int \psii(\vcf{r'}) \: \chiv(\vcf{r'};\vcrg{\theta}) \:
	\sst{\psi}{G}(\vcf{r}-\vcf{r'}) \, \text{d}\vcf{r'},
	\label{eq:par-est-first-Born-sensi}
\end{equation}
where the modified potential 
\begin{equation}
	\chiv(\vcf{r};\vcrg{\theta}) = \pder{n(\vcf{r};\vcrg{\theta})}{\vcrg{\theta}} \:
		n(\vcf{r}; \vcrg{\theta}).
	\label{eq:par-est-first-Born-chiv-def}
\end{equation}
Here, as in (\ref{eq:par-est-thin-film-sensi-transp}), the Sensitivity Field is produced by the same illumination as the original scattered field but with a modified scattering potential. 

One alternative to the Born series is the Rytov perturbation expansion, which starts by the transformation $\psis=\ee{\phis}$. To a first-order approximation then the scattered field is obtained as 
\begin{align}
	\psi(\vcf{r}) & \approx \psii(\vcf{r}) \: \expb{\breathe{0.5}{2.5} \phi_1(\vcf{r})}, 
	\qqwhere    \label{eq:par-est-Rytov-base} \\
	\phi_1(\vcf{r}; \vcrg{\theta}) & = \int \frac{\psii(\vcf{r'})}{\psii(\vcf{r})} \: 
	v(\vcf{r'}; \vcrg{\theta}) \: \psiG(\vcf{r}-\vcf{r'} )\: \text{d}\vcf{r'}.
	\label{eq:par-est-Rytov-integral} 
\end{align}
This result is known as the Rytov approximation or Rytov integral. Consequently, its Sensitivity Field is
\begin{align}
	\chis(\vcf{r};\vcrg{\theta}) & = \chii(\vcf{r}; \vcrg{\theta}) \: \expb{\breathe{0.5}{2.5} \phi_1(\vcf{r})}, 
	\qqwhere    \label{eq:par-est-Rytov-sensi-base} \\
	\chii(\vcf{r}; \vcrg{\theta}) & = \int \frac{\psii(\vcf{r'})}{\psii(\vcf{r})} \: 
	\chiv(\vcf{r'}; \vcrg{\theta}) \: \psiG(\vcf{r}-\vcf{r'} )\: \text{d}\vcf{r'}
	\label{eq:par-est-Rytov-sensi} 
\end{align}
is the sensitivity illumination field and $\chiv$ is again given by (\ref{eq:par-est-first-Born-chiv-def}). In this case, as in 
(\ref{eq:par-est-thin-film-sensi-illum}), the Sensitivity Field results from a modified illumination acting on the same scattering potential as in the original scattered field.

\subsection{Multiple scattering via the Lippmann-Schwinger equation \label{sec:par-est-lipp-schw} }

The Lippmann-Schwinger integral equation captures multiple scattering orders and is formally equivalent with, albeit usually better behaved numerically than the Born series. The scattered field is obtained from
\begin{equation}
	\psis(\vcf{r}) = \psii(\vcf{r}) + \int \psis(\vcf{r'}) \, v(\vcf{r'}) \, 
	\psiG(\vcf{r}-\vcf{r'}) \, \text{d}\vcf{r'}.
	\label{eq:par-est-lipp-schw}
\end{equation}
The scattering potential $v(\vcf{r})$ is the same as in (\ref{eq:par-est-scattering-potential}).

Taking gradients on both sides, the Sensitivity Field for the Lippmann-Schwinger integral equation is found to satisfy
\begin{equation}
	\chis(\vcf{r};\vcrg{\theta}) = \chii(\vcf{r}) + 
	\int \chis(\vcf{r'};\vcrg{\theta}) \, v(\vcf{r'};\vcrg{\theta}) \, 
	\psiG(\vcf{r}-\vcf{r'}) \, \text{d}\vcf{r'},
	\label{eq:par-est-lipp-schw-sensi}
\end{equation}
where the Sensitivity illumination field is
\begin{equation}
	\chii(\vcf{r};\vcrg{\theta}) = \int \psis(\vcf{r'};\vcrg{\theta}) \: \chiv(\vcf{r}; \vcrg{\theta}) \:
	\psiG(\vcf{r}-\vcf{r'}) \, \text{d}\vcf{r'}.
	\label{eq:par-est-lipp-schw-sensi-illum}
\end{equation}
Once again one arrives at a situation like (\ref{eq:par-est-thin-film-chii-n-d}) and (\ref{eq:par-est-Rytov-sensi}), where the Sensitivity Field is scattered by the same potential as the original scattered field with an illumination that results from the original scattered field modulated by the gradient of the scattering potential.

\subsection{Multiple scattering via the Beam Propagation Method \label{sec:par-est-bpm} }

If back scattering may be neglected, then a popular alternative is the Beam Propagation Method. In \cite{inv:pang-unified} it was shown that the Beam Propagation Method approximates the Lippmann-Schwinger equation solution well for weakly scattering objects, and the error grows as the fourth power of the maximum scattering angle, \ie\ the spatial bandwidth of the scatterer. 

In BPM, the object is divided into multiple slices oriented perpendicular to the directional axis $z$. Each slice is treated like a thin transparency, and subsequently the field is forward-propagated as if through free space to the next slice. That is, 
\begin{multline}
	\psi(x,y,z_{l+1}) = \int \psi(x',y',z_l) \, 
	\expb{  \breathe{0.5}{2.5}  ik\, \varepsilon(x',y';z_l)} \\
	\psiG(x'-x,y'-y,z'-z) \, \dx'\dy'.
	\label{eq:BPM}
\end{multline}
The recursion is initialized with 
\begin{equation}
	\psi(x,y,z_1) = \psii(x,y,z_1),
	\label{eq:BPM-ini}
\end{equation}
the incident illumination. Since the field from a given slice can reach the next slice at a considerably large angle, usually the ideal spherical wave kernel (\ref{eq:par-est-sphwa-kern}) is used in BPM.  

The sensitivity analysis on (\ref{eq:BPM}) was carried out in \missing{Ref Kamilov} leading to an elegant analogy between the sensitivity equation (\ref{eq:par-est-loss-fun-sensi}) for BPM and backpropagation in neural networks. Even though these papers considered only the pixel-oriented spatial imaging parameterization, \#1 in Section~\ref{sec:par-est-para-exs}, it is straightforward to generalize them to other parameterizations. 

\section{Conclusions \label{sec:par-est-concl} }

There are two main messages from this paper. The first is that treating the quantitative phase imaging problem as one of parameter estimation allows some interesting generealizations that are less ``anthropomorphic.'' Thus, perhaps accuracy improvements and efficiency savings may result. The parametric approach is inclusive of the traditional spatial imaging way of thinking, as the reader has seen in case \#1 of Section~\ref{sec:par-est-para-exs}. 

The second main point is the Sensitivity Field. It results in straightforward fashion from taking gradients on the loss function, eqs.~\ref{eq:par-est-loss-fun-def}-\ref{eq;par-est-sensi-fun-def}, with respect to the sought parameters. Superficially, thanks to automatic differentiation it may appear that there is no reason to carry out all the derivations of Sensitivity Fields in Section~\ref{sec:par-est-sensi-mod}. I beg to disagree. The results reveal an interesting intuition: the Sensitivity Fields are either produced by the same given scattering potential but with a modified Sensitivity Illumination field that is computed from the original scattered field, as in eqs.~\ref{eq:par-est-thin-film-sensi-illum}, \ref{eq:par-est-Rytov-sensi-base}, \ref{eq:par-est-lipp-schw-sensi}; or by the same illumination field but instead with a scattering potential that is the gradient of the original potential, as in eqs.~\ref{eq:par-est-thin-film-sensi-transp} and \ref{eq:par-est-first-Born-sensi}. Such analogies are well known in the community of feedback control\ \cite{book:pontryagin-optimal,dyn:cao03-adjoint-sensi} and have led to interesting innovations. \cite{dyn:chen19-neural-odes} 

\subsection*{Acknowledgments \label{sec:par-est-ack} }

This research was funded by the U. S. Air Force Office of Scientific Research through the Multi-University Research Initiative (MURI) ``Searching for what's new: the systematic development of dynamic x-ray microscopy,'' grant No. \linebreak FA9550-23-1-0284; and by Singapore's National Research Foundation through the Intra-Create thematic grant ``Retinal Analytics through Machine learning aiding Physics'' (RAMP), grant No. NRF2019-THE002-0006. The author is grateful to Steven G. Johnson of the MIT Mathematics Department; Chris Jacobsen of the Northwester University Physics Department; Shen Zuowei and Ji Hui of the National University of Singapore's Department of Mathematical Sciences; Fabian Br\"au, Bingyao Tan, Michael Girard, Leopold Schmetterer and Aung Tin of the Singapore Eye Research Institute; Sandip Mondal of SMART; and Qihang Zhang of Tsinghua University for helpful discussions.


\pagebreak

\appendix

\section*{Appendix: Derivation of the thin film approximation \label{sec:thin-film} }

Starting with the space-variant Helmholtz equation (\ref{eq:sca-wave-eq}), we elect a directional optical beam along the $z$ axis as
\begin{equation}
	\psi(\vcf{r}) = \tilpsi(\vcf{r}) \: \ee{ikz}.
	\label{eq:tilde-psi-def}
\end{equation}
Here, $\tilpsi(\vcf{r})$ is a slow-varying modulation , \ie
\begin{equation}
	\left| \pder{\tilpsi}{w}  \right| \ll k \quad \text{for\ } w=x,y,z.
	\label{eq:tilde-psi-conds}
\end{equation}
Substituting (\ref{eq:tilde-psi-def}) into (\ref{eq:sca-wave-eq}) we obtain
\begin{equation}
	\pderh{\tilpsi}{x}{2} + \pderh{\tilpsi}{y}{2} + \pderh{\tilpsi}{z}{2} + 
	2ik \, \pder{\tilpsi}{z} + k^2 \, 
	\left[ \breathe{0.5}{2.5} n^2(\vcf{r}) - 1 \right] \, \tilpsi(\vcf{r}) = 0.
	\label{eq:tilpsi-pde}
\end{equation}

We now make the strong assumption 
\begin{align}
	\left| \pderh{\tilpsi}{w}{2} \right|  &  \ll k \: \pder{\tilpsi}{z}, \quad \text{for\ }
	w=x,y,z; \qqand \label{eq:tilpsi-approx-del} \\
	\pder{n}{z} & = 0. \label{eq:n-no-z}
\end{align}
In Geometrical and Hamiltonian optics, condition (\ref{eq:n-no-z}) is also known as a ``guide medium.'' The remaining terms in (\ref{eq:tilpsi-pde}) then have the solution
\begin{equation}
\tilpsi(x,y,z) = \tilpsi_0(x,y) \, \expb{ik \, \frac{n^2(x,y)-1}{2} \, z},
\label{eq:tilpsi-thin-sq}
\end{equation}
where $\tilpsi_0(x,y) \equiv \tilpsi(x,y,0)$ is the incident field. To bring this into an even more recognizable form, we further assume a weak refractive index modulation
\begin{equation}
	n(x,y) = 1 + \varepsilon(x,y), \qqwhere \left|\varepsilon(x,y)\right| \ll 1.
	\label{eq:n-weak}
\end{equation}
Then (\ref{eq:tilpsi-thin-sq}) becomes
\begin{equation}
	\tilpsi(x,y,z) = \tilpsi_0(x,y) \, 
	\expb{ \breathe{0.5}{2.5}  ik \, \varepsilon(x,y) \, z}.
	\label{eq:tilpsi-thin}
\end{equation}
In the exponent, we recognize the optical path difference
\begin{equation}
	\text{OPD}(x,y) = \varepsilon(x,y) \, z.
	\label{eq:OPL}
\end{equation}
Clearly, the thin dielectric object under these approximations imparts a {\em pure phase modulation} on the incident field. This is a key consequence and justification for the term ``phase object.''

To summarize, the phase object or thin film approximation (\ref{eq:tilpsi-thin}) relies on all of the following assumptions to be true: 
\romlist{thin-conds}
\item The refractive index is independent of the axial variable $z$, according to (\ref{eq:n-no-z}). 
\item The refractive index modulation is small relative to the surrounding medium, according to (\ref{eq:n-weak}).
\item The following three conditions that are necessary for (\ref{eq:tilpsi-thin}) to be consistent with~(\ref{eq:tilpsi-approx-del}):
\begin{align}
	\left| \pderh{\tilpsi_0(x,y)}{w}{2}\right| & \ll 2k^2\varepsilon(x,y), 
	\label{eq:tilpsi-conds-1} \\
	\left| \pder{\tilpsi_0}{x}(x,y) \pder{\varepsilon(x,y)}{w} z\right| & \ll k \, \varepsilon(x,y), \label{eq:tilpsi-conds-2}  \\
	\left| \tilpsi_0(x,y) \pderh{\varepsilon(x,y)}{w}{2} z\right| & \ll 2k \, \varepsilon(x,y),  \label{eq:tilpsi-conds-3}
\end{align}
\end{list}
where the spatial derivatives are with respect to $w=x,y$. Without going into a full analysis of these expressions, one might just point out that as the depth $z$ increases, so does the potential of (\ref{eq:tilpsi-conds-1}-\ref{eq:tilpsi-conds-3}) becoming violated. The same holds for the lateral bandwidth of the incident illumination $\tilpsi_0(x,y)$ and the modulation $\varepsilon(x,y)$.

Remarkably, biological cells seem to satisfy these smoothness conditions most of the time. This is probably why QPI has been fairly succesful in visualizing them and identifying correlates to biological functions. On the other hand, it is  also cutomary to apply the thin film approximation to certain artificial objects that seem to break it, e.g. if they include sharp edges. This is the case for almost all binary optics and 3D-printed dielectric objects, for example. Even more remarkably, the thin-film approximation even then seems to predict the  output of the overall optical system in good agreement with experiments.  

It is possible that such objects in their physical realization are not actually as sharp as their models assume. Moreover, usually the light scattered from the highly sloped edges is directed outside the aperture of the optical system. Its absence from the detected intensity would appear to cause no damage other than smoothening; which may just be ignored or corrected by use of prior knowledge, \ie\ an edge-enhancing regularizer such as Total Variation (TV).


\pagebreak

\bibliographystyle{osajnl}
\bibliography{abbrev,books,Fneural,inverse,3di,optics,Fvh01,dynsys}

%

\end{document}